\documentclass[sigconf,numbers]{acmart}
\AtBeginDocument{%
  }

\setcopyright{acmlicensed}
\copyrightyear{2027}
\acmYear{2027}
\acmDOI{XXXXXXX.XXXXXXX}
\acmConference[SIGCSE ' 27]{Special Interest Group for Computer Science Education}{Feb 17--20, 2027}{Sacremento, CA}
\acmISBN{978-1-4503-XXXX-X/2027/02}

\usepackage{xcolor}
\usepackage{array}
\usepackage{graphicx}
\usepackage{subcaption}
\usepackage{mdframed}
\usepackage{xspace}
\usepackage{natbib}
\usepackage{tabularx}
\usepackage{todonotes}

\begin{document}

\title{Decomposing the Doer Effect in Programming Practice: Code Writing Stands Out Among Active Practice}

\author{Arun-Balajiee Lekshmi-Narayanan}
\email{arl122@pitt.edu}
\orcid{0000-0002-7735-5008}
\affiliation{%
  \institution{University of Pittsburgh}
  \city{Pittsburgh}
  \state{Pennsylvania}
  \country{USA}
}

\author{Gillian Gold}
\email{gilliang@andrew.cmu.edu}
\orcid{0009-0007-9220-8915}
\affiliation{%
  \institution{Carnegie Mellon University}
  \city{Pittsburgh}
  \state{Pennsylvania}
  \country{USA}
}
\author{Jordan Barria-Pineda}
\email{jordan.barria@udp.cl}
\orcid{anon}
\affiliation{%
  \institution{Universidad Diego Portales}
  \city{Santiago}
  \state{}
  \country{Chile}
}
\author{Quinn K Wolter}
\email{qkw3@pitt.edu}
\orcid{anon}
\affiliation{%
  \institution{University of Pittsburgh}
  \city{Pittsburgh}
  \state{Pennsylvania}
  \country{USA}
}
\author{Peter Brusilovsky}
\email{peterb@pitt.edu}
\orcid{0000-0002-1902-1464}
\affiliation{%
  \institution{University of Pittsburgh}
  \city{Pittsburgh}
  \state{Pennsylvania}
  \country{USA}
}
\author{Paulo Carvalho}
\email{pcarvalh@cs.cmu.edu}
\orcid{0000-0002-0449-3733}
\affiliation{%
  \institution{Carnegie Mellon University}
  \city{Pittsburgh}
  \state{Pennsylvania}
  \country{USA}
}








\renewcommand{\shortauthors}{Anon et al.}

\begin{abstract}
The "doer effect" suggests that actively doing practice activities is more strongly associated with learning outcomes than passively viewing content. In the doer effect literature, "doing" refers specifically to active practice. However, this categorization treats different forms of active practice as equivalent, leaving open whether some types of active practice are more effective than others. In this paper, we investigate whether the doer effect extends to computer science instruction and whether some forms of doing stand out compared to other forms. We analyze log data from 334 students across 11 semesters of introductory and intermediate Java who used an interactive practice system with five content types: Code Writing, Code Tracing, Code Completion, Code Visualizations, and Code Explanations. Consistent with prior doer effect work, we find that active practice activities were associated with 3.2 times better learning outcomes than passive activities. Interestingly, among the active practice, code writing was the most strongly associated with improved posttest performance, while no other activity type showed a comparable association. These results highlight the importance of challenging, feedback-supported practice activities, such as code writing problems.
\end{abstract}



\received{}
\received[revised]{}
\received[accepted]{}

\maketitle

\section{Introduction}
In a modern computer science (CS) educational learning system, students engage in a range of practice activities - exploring worked examples, assessing their knowledge with multiple choice questions, and solving different kinds of programming problems. The ``doer effect''~\cite{koedinger2015learning,carvalho2017is} describes a robust pattern in online courses: students who engage more with active practice activities show stronger learning gains than students who engage more with passive content — a finding with clear implications for instructional design. In programming education specifically, practice activities are heterogeneous: students may write code, trace its execution, assemble code from fragments, view animated visualizations of program execution, or read annotated code examples. While these can be broadly described as active and passive activities, some require students to produce a response that the system evaluates (active activities), while others present content that students explore without producing a response (passive activities), even within these broad categories there is great variability. For example, writing code and assembling code fragments, while both active learning activities require different types of engagement and likely cognitive processes.

With this in mind, there are two open questions for programming education practice systems specifically. First, the doer effect's active-vs-passive comparison has typically been tested against passive content consisting of text reading or basic instructional videos. In a multi-content integrated practice system, the passive content itself can be substantively interactive learning content — stepping through code execution and offering click-to-reveal line-by-line explanations (code explanations) — rather than text reading. It is not yet established whether the active-vs-passive distinction holds when the passive side is also interactive learning content. Second, both the active and passive categories contain multiple types — within active practice, students can write code, trace code execution, or assemble code from fragments; within passive content, students can view animated execution stepping or study annotated code examples. The doer effect framework's aggregate active-vs-passive comparison does not distinguish among these types. We investigated whether different types of practice activities contributed differently to learning.
 
We asked two research questions:
 
\begin{quote}
\textbf{RQ1.} Does the active-vs-passive doer effect pattern hold in a multi-content practice system where the passive materials are themselves substantively interactive learning content?
 
\textbf{RQ2.} Within the "doing" activities, do some types of practice stand out comparatively to other types?
\end{quote}
 
We investigated these questions using a dataset from introductory and intermediate Java courses where students used Mastery Grids, a practice system that integrated five activity types in a unified interface: \textbf{Code Writing, Code Tracing, Code Explanations, Code Visualizations, and Code Completions}. We addressed the questions through two regression models on log-derived data: an aggregate doer effect addressing RQ1, and the decomposition by activity type addressing RQ2.
 
Our analysis found that the doer effect pattern reported in prior work~\cite{koedinger2015learning,carvalho2017is} was also present in the Mastery Grids dataset available through DataShop~\cite{stamper2011human}. Within active activities, code writing emerged as the activity type most strongly associated with posttest performance in Model 2. Coefficients on the other active types (Code Tracing, Code Completion) and the two passive types were not statistcally significant in Model 2. We interpreted these findings as suggesting that corrective feedback, along with the active-vs-passive engagement, may explain the doer effect.

\section{Related Work}
\subsection{The Doer Effect in Online Courses}
 
The doer effect refers to the finding that completing practice activities is associated with greater learning gains than completing reading or viewing activities. It has been observed across multiple online deployments of the Open Learning Initiative (OLI) platform~\cite{koedinger2015learning,carvalho2017is}. For example, Carvalho et al.~\cite{carvalho2017is} reported that practice activities were associated with 2.4--3.6$\times$ better learning outcomes compared to reading, with the relationship robust across course content (e.g., psychology, computing) and delivery format (MOOC, blended). Subsequent work has examined whether the doer effect depends on the diversity of practice activities; Carvalho et al.~\cite{carvalho2022varied} found that completing many unique practice activities was associated with learning while repeating the same activities was not. Our analysis treats all student attempts as engagement with the activity, without distinguishing unique from repeated attempts; we discuss this measurement choice further in Section~\ref{sec:limitations}.
 
Two features of the doer effect literature motivate our investigation. First, the passive activities in prior doer effect comparisons are reading or video-watching tasks~\cite{koedinger2015learning,carvalho2017is} — the doer effect has not been tested against passive materials that are also interactive learning content. Second, the doer effect framework treats all "doing" as a single category. Whether different forms of doing — Code Writing, Code Completion, or Code Tracing — are equivalent practice modalities, or whether some are more strongly associated with learning outcomes than others, is not yet established within the doer effect framework. Our analysis addresses both of these gaps.
 
\subsection{Smart Content in Programming Practice}\label{sec:SLC}

Over the last 30 years, Computer Science researchers and practitioners have developed multiple types of interactive learning content that go beyond text, video, and multiple choice questions in introducing and assessing domain knowledge in the learning content. Frequently called "smart learning content''~\cite{brusilovsky2014increasing}, these types of content engage students in various learning activities, offer students a sense of control, and provide extended feedback. 
Two popular examples of ``smart content'' for programming are program visualizations~\cite{sorva2013review} and code-writing problems with automatic assessment~\cite{brusilovsky2005preface}, however the  review of the field recognized a wide variety of ``smart content types''~\cite{brusilovsky2014increasing}. Smart Learning Content (SLC) could be used in both assessment and practice modes. In the latter case, the students are encouraged to explore instructor-provided SLC to fill the gaps in their knowledge and check their understanding. Nowadays, many research teams offer platforms for sharing reusable SLC items such as ACOS server~\cite{sirkia2017acos} and CodeHarbor~\cite{staubitz2017towards}. SPLICE Infrastructure project collected thousands of SLC items in its catalog and offered a taxonomy of smart content types~\cite{SPLICE-Taxonomy}.

\subsubsection{Active vs Passive Learning Content}
The SPLICE taxonomy of SLC types splits all types of SLC in two main branches - active content (exercises) and passive content (presentations)~\cite{SPLICE-Taxonomy}. Active content requires students to engage in problem-solving and
to produce a response that the system evaluates as correct or incorrect. Passive content doesn't require a response and provides no correctness feedback — students explore it at their own pace.

This split is grounded in multi-year educational technology research. Chi and Wylie's ICAP framework~\cite{chi2014icap} characterizes learner engagement along four modes --- interactive, constructive, active, and passive --- with progressively richer cognitive engagement associated with stronger learning outcomes. The SPLICE active/passive division approximately corresponds to whether content elicits constructive output from the student or presents content for passive uptake. Among passive content types, worked examples are a particularly well-studied case: Chi et al.~\cite{chi1989selfexplanations} found that students who spontaneously generated self-explanations while studying worked examples learned more than students who studied them passively, and the worked-example effect~\cite{renkl2014toward} has been replicated extensively as an instructional tool for novice learners.

In this paper, we adopt active/passive decomposition principle and focus on exploring several active SLC types to compare educational effectiveness of several SLC types within the active category.

\subsubsection{Construction vs. Tracing in Programming Practice}
  
Among active learning content types explored in the paper, two approaches are represented that likely involve different cognitive processes: code construction (which includes code writing and completion) and code tracing. Construction requires composing or assembling constructs to achieve a goal state; tracing requires simulating execution to predict an outcome state. Sudol-DeLyser et al.~\cite{sudol2012code} argue that writing and tracing are theoretically distinct skills, with tracing positioned as a precursor to construction. An extensive body of empirical work in CS education has documented this dichotomy and the relationship between tracing and writing skills: Lopez et al.~\cite{lopez2008relationships} demonstrated a hierarchy of programming skills with tracing performance correlating with and predicting writing performance, a hierarchy replicated by Lister et al.~\cite{lister2009further} and further refined by Venables et al.~\cite{venables2009closer}. Tracing exercises ask students to predict execution outputs (for example, the value of a variable after a loop completes, or the console output of a code snippet). Sudol-DeLyser et al.\ emphasize the ``in-motion'' property of program execution (variables that change values across iterations, expressions evaluated step by step) as the key conceptual difficulty that tracing exercises address. Critically, they argue that the relationship between tracing and construction is asymmetric: students can trace successfully without being able to produce code, but reliable code production requires the underlying execution-simulation skill that tracing exercises.
 
This characterization of tracing and construction as distinct skills motivates our decomposition by activity type in our analyses. Within the ``doing'' category of the doer effect framework~\cite{koedinger2015learning},
Code writing and completion involve construction, while code tracing involves mental simulation. A decomposition by activity type lets us examine which of these distinct active activities is most strongly associated with posttest performance.
 
Prior work establishes that (a) the doer effect favors active over passive practice in online courses; (b) the decomposition principle from Hosseini et al.~\cite{hosseini2016animated} applies broadly to multi-content deployments; and (c) within active practice, construction and tracing involve different mental processes, with tracing positioned as a precursor to construction~\cite{sudol2012code}. Our contribution applies the decomposition principle to the active side of the active/passive division, showing that within Mastery Grids practice, Writing is the active type most strongly associated with the aggregate doer effect on a mixed-format posttest.

\section{Method}
 
\subsection{Dataset}
 
In this study, we use a dataset from an online practice system, Mastery Grids~\cite{loboda2014mastery}, collected over multiple semesters of classroom studies in introductory and intermediate Java programming courses. Mastery Grids datasets can be found in DataShop~\cite{stamper2011human}. Mastery Grids offers a unified interface to practice with multiple SLC types. It visualizes student topic-by-topic progress across the course curriculum and allows them to select practice activities by topic and type. Use of the system is encouraged but not required for course credit; students who do not engage with the system in any given week incur no grade penalty. 
In these studies, Mastery Grids provided access to five types of SLC activities collected by SPLICE project (cssplice.org) from four different content providers. Code writing activities were provided by the PCRS system~\cite{zingaro2013facilitating}, code tracing activities were provided by the QuizJet system~\cite{hsiao2010guiding}, code completion and code explanation activities came from the PCEX system~\cite{hosseini2018pcex}, and code visualization activities came from Jsvee~\cite{sirkia2018jsvee}.
We categorize activities in the dataset in Table~\ref{tab:taxonomy} along active/passive and construction/tracing dimensions introduced in Section~\ref{sec:SLC}.

\begin{table}[t]
\caption{Activity taxonomy in Mastery Grids. The active/passive distinction reflects whether the activity provides corrective feedback. We further distinguish active (writing, completion, tracing) from passive (visualizations, explanations) following~\cite{sudol2012code}.}
\label{tab:taxonomy}
\centering\small
\begin{tabular}{p{2.9cm}p{4.7cm}}
\toprule
Category & What students do \\
\midrule
Code \textbf{Writing}~\cite{zingaro2013facilitating}:  
Active / Construction& Write Java code from scratch, evaluated by autograder \\
Code \textbf{Completion:}~\cite{hosseini2018pcex} Active / Construction & Parsons-like code construction with explanations after each attempt  \\
Code \textbf{Tracing:}~\cite{hsiao2010guiding} Active / Tracing & Enter the expected output of a given code snippet; receives correct/incorrect feedback without explanation \\
Code \textbf{Visualization:}~\cite{sirkia2018jsvee} Passive / Tracing & View visualizations of code execution showing the stack and console output line-by-line\\
Code \textbf{Explanation:}~\cite{hosseini2018pcex} Passive / Construction & Read line-by-line explanations of code construction examples (no right/wrong feedback) \\
\bottomrule
\end{tabular}
\end{table}

\begin{figure}[h]
\centering
\includegraphics[width=\columnwidth]{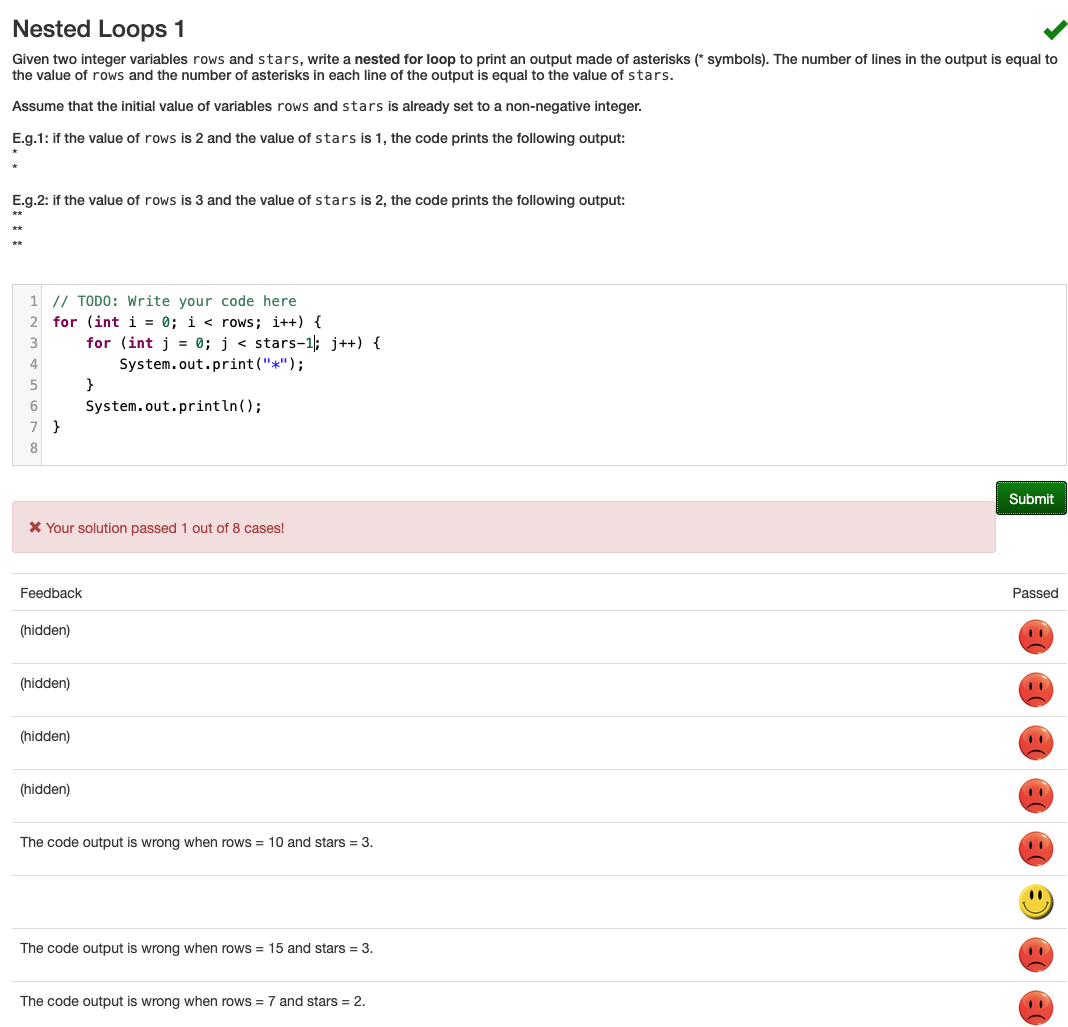}
\caption{An example of an active content type~\cite{zingaro2013facilitating}-- code writing, which provides corrective feedback on student responses.}
\label{fig:active-activities}
\end{figure}
 
\begin{figure}[h]
\centering
\includegraphics[width=\columnwidth]{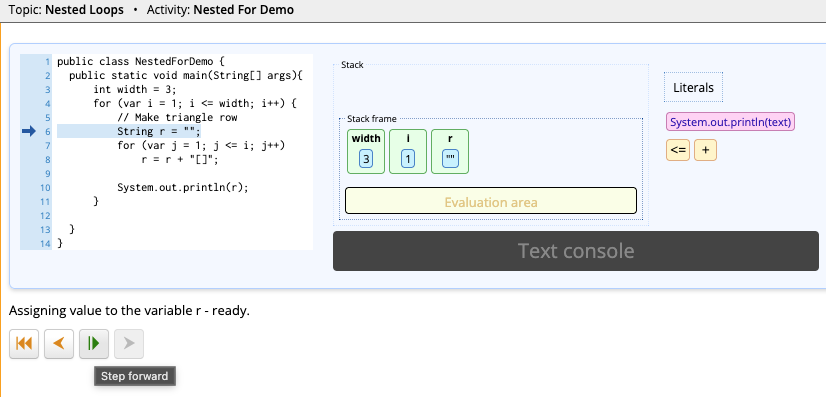}
\caption{An example of a passive activity type  -- code visualization~\cite{sirkia2018jsvee}, which explores content at their own pace with no corrective feedback.}
\label{fig:passive-activities}
\end{figure}

Table~\ref{tab:inventory} summarizes the practice content available to students. The platform contained 362 practice problems and 20 problems used for the pretest and posttest and 143 examples. Problem activities were the most numerous (130 code completion, 117 code tracing, and 115 code writing problems); among passive content, code construction examples with explanations (100) substantially outnumbered code visualization examples (43).
 
\begin{table}[h]
\caption{Content inventory in MasteryGrids. Practice-content counts
exclude items used in the 10-item pretest and 10-item posttest.
Internal platform acronyms are shown in parentheses for cross-reference
with Table~\ref{tab:taxonomy}.}
\label{tab:inventory}
\centering\small
\begin{tabular}{lr}
\toprule
Content type & Items available \\
\midrule
Completion Problems & 130 \\
Tracing Problems & 117 \\
Writing Problems & 115 \\
Visualization Examples & 43 \\
Explanation Examples & 100 \\
\midrule
Problems & 362 \\
Examples & 143 \\
\midrule
Pretest Questions& 10 \\
Posttest Questions& 10 \\
\bottomrule
\end{tabular}
\end{table}

\subsection{Participants and Sample Construction}
 
Data were collected from students enrolled in 11 introductory and intermediate Java programming courses taught in the years 2021 through 2025 at large US based universities. All courses were taught by instructors from the research team and data was collected with participant content. Course topics covered the full introductory Java curriculum including the following topics: \texttt{Variables and Operators, Strings, Boolean Expressions, If-Else, While Loops, For Loops, Nested Loops, Objects and Classess, Arrays, ArrayLists, Exception Handling, File Processing, Inheritance, Recursion, Searching, Sorting, and Advanced OOP}.

\begin{table}[h!]
    \centering
    
    \caption{Student Interactions with Smart Content on Mastery Grids -- extreme values were capped between 5th and 95th percentiles. in parenthesis the values are standard deviations}
    \label{tab:dataset_summary}
    \begin{tabular}{|c|c|}
    \hline
        N (overall) & 334 \\ \hline
        N (active) & 204  \\ \hline
        N (completed pretest and posttest) & 96 \\\hline  
        Average Pretest&  0.54 (0.32) \\ \hline
        Average Posttest& 0.83 (0.25) \\\hline
        
       Average Completed Problems&  109.38 (68.86) \\ \hline
       Average Example Visits& 38.26 (30.12) \\ \hline
       
       Average Writing Problems& 34.73 (1) \\ \hline
       Average Completion Problems& 26.25  (1) \\ \hline
       Average Tracing Problems& 39.43 (1)\\ \hline
       Average Explanation Examples& 17.01 (0) \\ \hline
       Average Visualizations Examples& 19.36 (0) \\\hline
         
    \end{tabular}
\end{table}

There was substantial attrition from total enrollment to the analytic sample, which reflects the requirements of completing both pretest and posttest, along with the voluntary nature of the Mastery Grids deployment. A comparable engagement-rate pattern was reported by Brusilovsky et al.~\cite{brusilovsky2018integrated} in a similar voluntary multi-content deployment, where approximately half of the enrolled students did not engage with the practice system at all. The analytic-sample size we report here is therefore representative of voluntary-deployment studies of integrated practice systems rather than a feature specific to our study.
 
\subsection{Outcome and Predictor Variables}
 
\textbf{Pretest score} measures students' prior content knowledge at the start of the term and is computed as
\texttt{corrects\_pre / total\_pre}, the fraction of correctly answered items on the 10-item multiple-choice pretest.
\textbf{Posttest score} is the primary outcome variable and is computed as \texttt{corrects\_post / total\_post}, the fraction of correctly answered items on the 10-item multiple-choice posttest. The posttest contains a mix of tracing items (predict the output of a code snippet) and construction items (write or complete a Java function), in approximately equal proportion.
\textbf{Activity counts} are the predictor variables of interest. For each student, we count all attempts across the practice period for each of the five activity types defined in Table~\ref{tab:taxonomy}.
 
\subsection{Analysis}
\label{sec:method-strategy}
 
We use ordinary least squares regression (the \texttt{lm} function in R) with standardized predictors. Our analytical strategy fits two models that we refer to as Model 1 and Model 2 throughout the paper.
 
Model 1 (aggregate doer effect) addresses RQ1; it estimates the aggregate doer effect relationship in the cross-sectional data set (one row per student) using summed counts of problem attempts and sample visits, in the active-vs. passive specification used in previous doer effect studies of voluntary online courses~\cite{koedinger2015learning,carvalho2017is}.
 
Model 2 (type decomposition) addresses RQ2; it re-fits the cross-sectional model with the active and passive categories decomposed by type — all five activity-type counts entered as separate predictors instead of two aggregate categories. 

To assess whether class-level clustering affected the estimates, we additionally fit mixed-effects versions of Models 1 and 2 with a random intercept for class (the \texttt{lmer} function from the \texttt{lme4} package, with Satterthwaite-approximated p-values via \texttt{lmerTest}). Class-level variance components and intraclass correlations are reported alongside the cross-sectional results.

Throughout, all predictors are standardized to $z$-scores. Significance markers follow convention: $^{***}\,p < .001$, $^{**}\,p < .01$, $^{*}\,p < .05$, $^{.}\,p < .10$.

\section{Results}~\label{sec:results}

\subsection{RQ1: Modeling the Aggregate Doer Effect}
\label{sec:r-doer}
 
Model 1 used the active-vs-passive specification used in prior OLI doer effect studies~\cite{koedinger2015learning,carvalho2017is}, and all predictors were z-scored. We regressed posttest score on pretest score, along with the total counts: all problem attempts (sum of attempts across the three active types — writing, tracing, completion) and all example visits (sum of visits across the two passive types — visualization, explanation).
 
\begin{table}[h]
\caption{Model 1: aggregate doer effect model (cross-sectional).
$n = 111$, $R^2 = 0.28$, adjusted $R^2 = 0.26$,
$F(3, 107) = 14.19$, $p = 7.50 \times 10^{-8}$.
Significance: $^{***}\,p<.001$, $^{**}\,p<.01$, $^{*}\,p<.05$,
$^{.}\,p<.10$.}
\label{tab:mdl-doer}
\centering\small
\begin{tabular}{lrrr}
\toprule
Predictor & $\beta$ & SE & $t$ \\
\midrule
Pretest & 0.513 & 0.086 & $5.965^{***}$ \\
All Problem Attempts & \textbf{0.351} & 0.095 & $3.713^{***}$ \\
All Example Visits & $-0.112$ & 0.095 & $-1.181$ \\
\bottomrule
\end{tabular}
\end{table}
 
The active coefficient ($\beta = 0.351$, $t = 3.71$, $p < .001$) was approximately 3.2 times stronger than the passive coefficient ($|\beta| = 0.112$, $t = -1.18$, $p = 0.240$). The pattern was consistent with the active-versus-passive results reported in prior doer effect work~\cite{carvalho2017is}. This replicated the active-vs-passive pattern of the doer effect.

The mixed-effects version of Model 1 with a random intercept for class returned a singular fit: the class-level intercept variance was estimated as zero ($\sigma^2_{\text{class}} = 0.000$, ICC $= 0$, $n_{\text{groups}} = 7$). With class-level variance estimated as zero, the mixed-effects model is algebraically equivalent to OLS, and the fixed-effect estimates and Satterthwaite-approximated p-values were identical to the cross-sectional values reported above. There was no detectable class-level clustering in post-test outcomes after controlling for pretest and practice engagement.

\subsection{RQ2: Modeling Decomposition by Activity Type}
\label{sec:r-disagg}
 
Model 2 used all five activity-type counts as separate predictors, controlling for pretest, with all predictors z-scored. We regressed posttest score on pretest score plus five activity-type counts: writing, tracing, and completion (the three active types), and visualization and explanation (the two passive types).
 
\begin{table}[h]
\caption{Model 2: decomposition by activity type (cross-sectional).
$n = 96$, $R^2 = 0.32$, adjusted $R^2 = 0.28$,
$F(6, 89) = 7.01$, $p = 3.97 \times 10^{-6}$.
Significance: $^{***}\,p<.001$, $^{**}\,p<.01$, $^{*}\,p<.05$,
$^{.}\,p<.10$.}
\label{tab:mdl-disagg}
\centering\small
\begin{tabular}{lrrr}
\toprule
Predictor & $\beta$ & SE & $t$ \\
\midrule
Pretest & 0.443 & 0.091 & $4.864^{***}$ \\
Writing & \textbf{0.288} & 0.085 & $3.391^{**}$ \\
Tracing & $-0.107$ & 0.167 & $-0.638$ \\
Completion & $0.183$ & 0.138 & $1.322$ \\
Visualization & $-0.279$ & 0.278 & $-1.005$ \\
Explanation & $0.297$ & 0.248 & $1.200$ \\
\bottomrule
\end{tabular}
\end{table}
 
Students who completed more code writing problems also had higher posttest scores ($\beta = 0.288$, $t = 3.39$, $p = 0.00104$). Students who completed more tracing problems, more completion problems, or who viewed more  visualization or explanation examples did not have systematically different posttest scores. Ultimately, among the five activity types in Mastery Grids, code writing practice was most strongly associated with posttest outcomes.

The mixed-effects version of Model 2 (random intercept by class) also returned a singular fit, with class-level intercept variance estimated as zero ($\sigma^2_{\text{class}} = 0.000$, ICC $= 0$, $n_{\text{groups}} = 7$). The fixed-effect estimates and Satterthwaite-approximated p-values were identical to the cross-sectional Model 2 values. The within-active pattern — code writing as the only reliably associated active type — held under the mixed-effects specification.
 
\section{Discussion}
\subsection{The Advantage of Active Practice}
\label{sec:doer-discussion}
Our first finding extends the doer effect literature to a programming learning environment in which the passive activities are themselves interactive. Rather than consisting of videos or static text, the passive materials in Mastery Grids require students to step through animated program execution (code visualization) and reveal line-by-line explanations (code explanations). Despite this level of cognitive engagement, students who devoted relatively more effort to practice problems demonstrated larger learning gains than those who relied more heavily on passive materials. The magnitude of this active-versus-passive difference was also at the upper end of the 2.4--3.6$\times$ range reported by Carvalho et al. \cite{carvalho2017is} for OLI courses, suggesting that the doer effect generalizes beyond traditional instructional media to interactive programming environments.

At the same time, our findings suggest that the distinction between "doing" and "viewing" may not fully characterize why active practice is effective. Both the active and passive activities required students to engage  with code, yet only the active activities were associated with greater learning. One plausible explanation is the role of corrective feedback. Active activities provide immediate evaluation of students' responses, allowing learners to identify misconceptions, revise their solutions, and iteratively refine their understanding. Feedback has long been recognized as an essential component of programming instruction and intelligent tutoring systems \cite{keuning2018systematic, messer2024automated}, although evidence regarding the magnitude of its effects remains mixed~\cite{van2015effects}. Our results suggest that practice with corrective feedback may be an important mechanism underlying the doer effect in programming practice, but experimental studies that independently manipulate feedback availability will be needed to determine the extent to which feedback, rather than activity type itself, drives these learning gains.

From a practical perspective, our findings reinforce the primary instructional implication of the doer effect: programming practice systems should encourage students to actively solve problems rather than passively consume instructional content. Even when passive materials are designed to promote some cognitive engagement, opportunities for learners to generate and test their own solutions appear to provide additional learning benefits.

\subsection{Not All Active Practice Is Equal}

Our second contribution refines the doer effect by examining whether all forms of active practice contribute equally to learning. The original doer effect framework treats active engagement as a single category. In contrast, Mastery Grids includes multiple forms of active practice — code writing, tracing, and code completion — that place different cognitive demands on learners. Our results suggest that these activities should not be considered interchangeable. Among the three activity types, only code writing was consistently associated with improved posttest performance after accounting for the other forms of engagement.

Several characteristics of code writing may explain this finding. Unlike tracing or completion, writing requires learners to retrieve programming knowledge, plan an entire solution, and generate code from scratch before receiving feedback~\cite{sudol2012code,lopez2008relationships}. These generative processes require learners to actively organize and apply their knowledge rather than recognize or interpret existing code, consistent with broader research demonstrating the benefits of retrieval-based and generative learning activities. After submission, automated feedback provides immediate information about correctness, allowing learners to debug their solutions and iteratively refine their mental models of program behavior \cite{watson2011corrective, keuning2018systematic, messer2024automated}. Our results cannot determine which of these features is primarily responsible for the observed association, but they suggest that the combination of generative practice and timely feedback may make code writing particularly effective.

Importantly, these findings should not be interpreted as evidence that tracing and completion lack instructional value. Both activities may support learning by reducing cognitive load, scaffolding novice programmers, or preparing students for more complex programming tasks. Rather, our results suggest that, when considered alongside writing activities, their independent contributions to posttest performance were not distinguishable in this dataset. Future work should investigate how these activity types can be sequenced most effectively—for example, whether tracing or completion problems are most beneficial as scaffolds that prepare learners for subsequent code writing rather than as substitutes for it.

From an instructional design perspective, our findings suggest that not all active learning activities should be emphasized equally. When resources are limited, investing in high-quality code writing problems with automated feedback may provide greater educational returns than expanding other forms of active practice. At the same time, understanding how writing, tracing, and completion can be optimally combined remains an important direction for future research.
 
\subsection{Limitations and Future Work}
\label{sec:limitations}
Although our work is consistent with prior literature, several limitations should be considered when interpreting these findings. First, our final analytic sample is relatively small ($n = 96$), reflecting the requirement that students complete both pretest and posttest assessments while engaging with the platform.  This reduced statistical power, particularly for models including multiple activity types and covariates, and may limit the detection of smaller effects. Our inclusion criterion also introduces selection bias: students who completed both pretest and posttest may differ systematically (in motivation, engagement, or academic preparation) from students who did not. Replication with larger and more diverse student populations would strengthen these findings.
 
Second, our analysis is observational, not experimental. Students chose which activities to engage with, and unobserved characteristics (motivation, prior programming experience beyond what the pretest measures, time available for the course) may drive both activity selection and posttest performance. Our pretest control addresses prior content knowledge but cannot control for self-efficacy, study time, or general academic preparation. Following the broader doer effect literature~\cite{carvalho2017is,carvalho2018analyzing}, we report associations without making causal claims. 
 
Third, Mastery Grids functioned as a supplemental practice system rather than the primary instructional environment. Students also learned through lectures, assignments, and other course activities that were not captured in our analyses. Consequently, the effects reported here represent the contribution of activity choices within the platform rather than the totality of students' learning experiences. Future work should examine whether these findings generalize to courses with different instructional models, programming languages, and learning platforms.
 
Fourth, our analyses treated every student attempt as an instance of engagement, including repeated attempts on the same activity. Carvalho et al.~\cite{carvalho2022varied} found that unique activity completion was more predictive of learning than repeated engagement, suggesting that these forms of practice may have different educational value. In future analyses, we should distinguish between unique and repeated attempts and investigate how repeated engagement influences learning across different activity types.

\section{Conclusion}
\label{sec:conclusion}
 We analyzed programming practice behaviors and posttest performance across 11 introductory and intermediate Java courses using Mastery Grids. Consistent with the doer effect \cite{koedinger2015learning}, students who engaged more frequently in active practice demonstrated substantially stronger learning gains than those who relied more heavily on passive instructional activities, extending prior findings to an interactive programming environment.

More importantly, our results suggest that active engagement should not be treated as a single, homogeneous construct. Among the types of active activity available in Mastery Grids, only code writing was consistently associated with improved posttest performance after accounting for other forms of practice. This finding suggests that the effectiveness of active learning depends not only on whether students are doing something, but also on the specific cognitive processes required by the activity. Code writing uniquely combines generative problem solving with immediate corrective feedback, making it a particularly valuable form of programming practice.

As programming education increasingly incorporates automated assessment, intelligent tutoring systems, and AI-supported practice environments, understanding which forms of active engagement most effectively promote learning will become increasingly important. Our findings suggest that investing in opportunities for students to construct complete solutions and receive timely feedback may have greater benefits than simply increasing the amount of interactive content, providing a more nuanced perspective on how active learning should be designed in modern programming practice systems.

\section{Acknowledgments}
Anonymized for blind review

\bibliographystyle{abbrv}
\bibliography{references.bib}


\end{document}